\documentclass[copyright,creativecommons]{eptcs}
\usepackage{graphicx} %
\usepackage[dvipsnames]{xcolor}
\usepackage{alltt}
\usepackage{framed}
\usepackage{subfig}
\usepackage{stmaryrd}
\usepackage{amsmath}

\usepackage{listings}
\newcommand{\sem}[1]{\llbracket #1 \rrbracket}

\lstdefinestyle{code}{
    language=RL,
    backgroundcolor=\color{white},   
    commentstyle=\color{gray},
    keywordstyle=\bfseries\color{violet},
    numberstyle=\tiny\color{gray},
    basicstyle=\ttfamily\footnotesize, 
    basewidth=0.55em, 
    breakatwhitespace=false,
    breaklines=false, 
    escapeinside={@}{@},
    captionpos=b, 
    lineskip=-2pt, 
    belowskip=0pt, 
    frame=lines, 
    keepspaces=true,                 
    numbers=left, 
    numbersep=5pt, 
    showspaces=false,                
    showstringspaces=false,
    showtabs=false,     
    xleftmargin=1.2em,
    framexleftmargin=1.em,
    framexrightmargin=-1.5em,
    tabsize=4,
    moredelim={[is][\color{BlueViolet}]{\#}{\#}} 
}

\lstdefinelanguage{RL}{
    morecomment=[l]{//}
}

\newcommand{\fontproperty}[1]{{\it{#1}}}                  %
\newcommand{\COMPS}[3]                                    %
                 {{\fontproperty{Comps}_{{#1},{#2}/{#3}}}}%

\newcommand{\fontsetlang}[1]{{\cal{#1}}}%

\newcommand{\setD}{\fontsetlang{D}}     %

\newcommand{\detd}[2]                   %
 {\mymath{{\setD}_{\mgdialect{#1}{#2}}}}%

\newcommand{\fontlang}[1]{{\mathit{#1}}}%

\newcommand{\Lv}{\fontlang{L}}          %
\newcommand{\Rv}{\fontlang{R}}          %
\newcommand{\mgdialect}[2]{{#1}|{#2}}   %
\newcommand{\sint}[2]                   %
                  {{\mymath{{#1}/{#2}}}}%
\newcommand{\nsint}[3]                  %
  {{\mymath{{\mgdialect{#1}{#2}}/{#3}}}}%
\newcommand{\scomp}[3]                  %
 {{\mymath{{{#1}\rightarrow{#2}}/{#3}}}}%

\newcommand{\intpgm}{\fontprogram{int}}                %
\newcommand{\inv}{\fontprogram{inv}} 
\newcommand{\spec}{\fontprogram{spec}}                    %
\newlength{\lindent}
\newenvironment{myindent}[2]%
         {\begin{list}{}{\setlength{\leftmargin}{#1}
               \setlength{\rightmargin}{#2}
               \setlength{\itemindent}{0cm}}%
            \item []}%
         {\end{list}}

\def\doframeit#1{\vbox{%
  \hrule height\fboxrule
    \hbox to \linewidth{%
      \vrule width\fboxrule \hss
      \vbox{\kern\fboxsep #1\kern\fboxsep }%
      \hss \vrule width\fboxrule }%
    \hrule height\fboxrule }}

\def\frameit{\smallskip \setbox0=\vbox\bgroup\advance \linewidth -2\fboxsep
  \advance \linewidth -2\fboxrule\hsize\linewidth
\strut \ignorespaces }

\def\endframeit{\ifhmode \par \nointerlineskip \fi \egroup
\doframeit{\box0}}
\newcommand{\mymath}[1]{\ensuremath{#1}}           %
\newcommand{\fontprogram}[1]{{\it{#1}}}                %

\newcommand{\pgm}{\fontprogram{p}}                     %
\newcommand{\pgmq}{\fontprogram{q}}                    %
 \newlength{\aLength}
 \newcommand{\aL}[1]                               
{\mymath{%
\stackrel{*\hspace*{0.15em}}{\Rightarrow}
\mbox{
\settowidth{\aLength}
{\scriptsize{#1}}%
\hspace*{-0.5\aLength}\hspace*{-0.9em}
\raisebox{-1.6ex}{\scriptsize{#1}}
\hspace*{-0.5\aLength}\hspace*{0.7em}
}
}}

\def\mynomath#1{\relax\ifmmode{\mathchoice%
{\hbox{#1}}%
{\hbox{#1}}%
{\hbox{\scriptsize{#1}}}%
{\hbox{\tiny{#1}}}%
}\else{\hbox{#1}}\fi}
\newcommand{\programmode}{%
        \small\tt%
        \catcode`\_=12 \catcode`\?=12 \catcode`\.=12 \catcode`\,=12
        \catcode`\;=12 \catcode`\:=12 \catcode`\@=12 \catcode`\~=12
        \obeyspaces\frenchspacing}%
\newcommand{\sbtt}{\mymath{[\![}\bgroup\programmode\sbtthack}
\newcommand{\sbtthack}[1]{\mbox{#1}\egroup\mymath{]\!]}}

\newcommand{\svbt}{\bgroup\programmode\svbthack}
\newcommand{\svbthack}[1]{\mbox{#1}\egroup}

\newcommand{\RTMinc}{\textsf{inc}}
\newcommand{\RTMdec}{\textsf{inc$^{-1}$}}
\newcommand{\ARL}{{ARL}}
\newcommand{\PEARL}{{PEARL}}
\newcommand{\RTM}{{\mathit{RTM}}}

\definecolor{LineGray}{gray}{0.75}
\definecolor{LightGray}{gray}{0.90}

\newcommand{\codeline}[0]{\color{LineGray}\rule[2.0pt]{0.88\textwidth}{0.4pt}} %

\usepackage{xcolor} %

\newcommand{\eg}{{e.g.}}

\newcommand{\ie}{{i.e.}}

\newcommand{\vs}{{vs.}}
\newcommand{\wrt}{{w.r.t.}}
\title{
Inversion by Partial Evaluation: \\
A Reversible Interpreter Experiment
}
\author{
Robert Gl\"uck \qquad Louis Marott Normann
\institute{DIKU, Department of Computer Science \\
University of Copenhagen, 
Denmark
}
\email{glueck@acm.org \quad lono@di.ku.dk}
}
\date{June 2024}

\def\titlerunning{Inversion by Partial Evaluation
}
\def\authorrunning{Robert Gl\"uck,
Louis Marott Normann}

\begin{document}
\maketitle

\begin{abstract}
A computational limit of combining partial evaluation and program inversion is investigated. Using a reversible Turing machine interpreter, we show that the first Futamura and inversion projections 
can produce not only functionally but also textually equivalent programs. 
The construction of the interpreter in a reversible {flowchart} language is shown in full.
Insights are provided on the practical interplay
between reversible interpreters, program inverters, and partial evaluators.
We conclude that both projections must
be included in the 
program transformation 
toolbox.

\end{abstract}

\section{Introduction}
\label{sec:introduction}

The \emph{Futamura projections} are a cornerstone of program transformation. They show, among other things, that programs can be translated by specializing interpreters~\cite{Futamura:71}. \emph{Partial evaluation} is a method for program specialization for which all three Futamura projections have been constructively realized~\cite{JSS:85}. 
Inspired by this success,
various transformations using partial evaluation have been investigated~\cite{ConselDanvy:93:POPL,DGT:96,JGS:93}, 
such as realistic compilation~\cite{SperberThiemann:96}, %
compiler generators~\cite{Thiemann:96,ThiemannSperber:96},
binding-time analysis~\cite{GlueckJoergensen:96:MBTA,GlueckNakashigeZoechling:95,Thiemann:97}, and
transformation by 
interpreter specialization~\cite{GHML:23,Jones:04:whatnot,SGT:96}.

\emph{Reversible computing} is an unconventional computing paradigm that,
like partial evaluation,
emerged 
in the 1970s~\cite{Bennett:IBM73} and has attracted %
renewed interest through applications in quantum-based and energy-efficient computing ({\eg}, \cite{DeVos:10:book,Krakovsky:21}).
The defining features 
are backward deterministic computation without information erasure and the 
invertibility of reversible programs~\cite{GlYo23:perspective}.
By contrast, mainstream languages are typically 
backward nondeterministic, 
and
program inversion
is a challenging 
transformation
problem
({\eg}, program calculation~\cite{BirdMu:04:BW,Dijkstra:78}, metacomputation~\cite{GlueckTurchin:90,Romanenko1991}, and inverter systems~\cite{GlueckKawabe:05:KEinv,Harrison:88,KawabeGlueck:05:LRS}).

Little is known about the interplay between 
specialization
and 
inversion in general and in reversible computing in particular; this is the primary motivation for the current study.
Reversible 
programming 
languages are particularly well suited for 
exploring the 
outer limits of 
such a non-standard 
combination,
as they allow the investigation of %
essentials 
in a clean scenario without the irreversible artifacts of mainstream languages. 

\begin{figure}[t] %
\noindent
\centering
\includegraphics[width=0.7\textwidth]{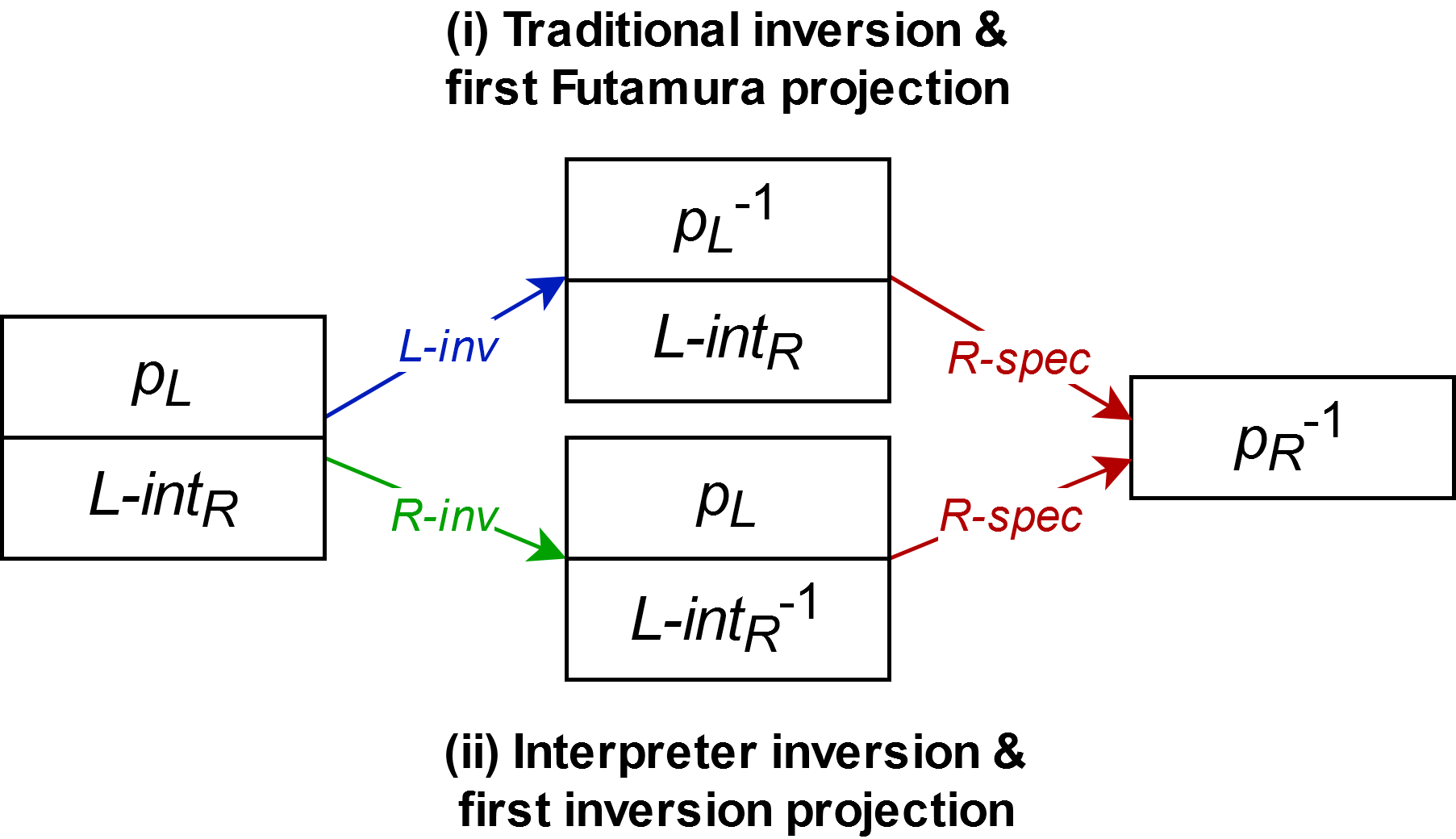}
\vspace{1ex}
\caption{Experimental setup: the first Futamura projection {\vs} the first inversion projection}
\label{fig:experimentsetup}
\end{figure}

Herein, we discuss one such 
limit: the inversion of 
programs by partial evaluation.
The experimental setup is illustrated in Fig.~\ref{fig:experimentsetup}. 
On the left is the starting point: a program $\pgm_\Lv$ written in $\Lv$ is interpreted by an $\Lv$-interpreter $\Lv$-$\intpgm_\Rv$ written in $\Rv$.
For notational simplicity, we assume {throughout the paper} that $\pgm_\Lv$ implements an \emph{injective function} such that a unique inverse function exists. 

\begin{enumerate}
\item[i.] The first (upper) path shows the inversion of $\pgm_\Lv$ into 
{the \emph{inverse program}} $\pgm^{-1}_\Lv$. This transformation is 
{performed} using an $\Lv$-inverter $\Lv$-$\inv$. 
This step is followed by the translation of $\pgm^{-1}_\Lv$ into $\pgm^{-1}_\Rv$ according to the \emph{first Futamura projection} by an $\Rv$-specializer $\Rv$-$\spec$. This path shows the \emph{sequential} transformation 
of %
$\pgm_\Lv$ into $\pgm^{-1}_\Rv$
by inversion and specialization.

\item[ii.] The second (lower) path shows the inversion of $\Lv$-$\intpgm_\Rv$ into 
{the \emph{inverse interpreter}} $\Lv$-$\intpgm^{-1}_\Rv$
using an $\Rv$-inverter $\Rv$-$\inv$. This step is followed by the inversion of $\pgm_\Lv$ into $\pgm^{-1}_\Rv$ according to the \emph{first inversion projection} by 
specializing $\Lv$-$\intpgm^{-1}_\Rv$.
The second step performs 
the \emph{simultaneous} inversion and translation of 
$\pgm_\Lv$ into $\pgm^{-1}_\Rv$.
{Note} that \emph{no} $\Lv$-inverter is required to invert the $\Lv$-program. %
{Inverting the $\Lv$-interpreter
by an $\Rv$-inverter 
is sufficient
and needs to be done just once.}
\end{enumerate}

The inverse programs resulting from both projections are guaranteed to be \emph{functionally equivalent}~\cite{AbrGlu02:URAF,Futamura:71};
this raises an interesting question: Can the programs have the \emph{same operational qualities}?
The theory of the Futamura and inversion projections reveals nothing about this {question}. 
To what extent must the residual programs differ operationally depending on the level at which the inversion is carried out?
Are there $\Lv$-interpreters for which the resulting programs are \emph{textually equivalent} ({\ie}, the
two
projections can be indistinguishable regarding their results)? 
Can 
the same $\Lv$-interpreter be the {starting point} for both projections?

This question can be briefly stated by omitting language annotations {and writing $\cong$ for the textual equivalence of the two programs
generated by specialization:}
\begin{eqnarray}\label{eq:challenge}
 \sem{\spec}(\intpgm, \pgm^{-1}) &\stackrel{\mbox{\footnotesize?}}{\cong}& \sem{\spec}(\intpgm^{-1}, \pgm)~.
\end{eqnarray}

Only recently have partial evaluation and reversible computing been combined 
{to investigate this question}~\cite[Sec.~5.2]{NormannGlueck:24}.
We performed the experiment shown in Fig.~\ref{fig:experimentsetup} in the following reversible setup: 
{$\Lv$-$\intpgm_\Rv$ is} the interpreter for \emph{reversible Turing machines} (RTMs) in Sec.~\ref{sec:RTM}, and {$\Rv$-$\spec$ is} the partial evaluator for the \emph{reversible flowchart language} {\ARL}~\cite{NormannGlueck:24}.
Thus, in Fig.~\ref{fig:experimentsetup}, $\Lv$ is the language of RTM-programs, and $\Rv$ is {\ARL}.
This paper reports new findings. 

The contributions of this study are as follows: 
\begin{itemize}
\item 
First 
demonstration 
that the Futamura and inversion projections can produce equivalent results.

\item 
A reversible Turing machine interpreter and inverter that enable successful projections.

\item 
Insights \hspace{-0.4pt}into \hspace{-0.4pt}the \hspace{-0.4pt}interplay \hspace{-0.4pt}between \hspace{-0.4pt}interpreters, \hspace{-0.4pt}inverters, \hspace{-0.4pt}and \hspace{-0.4pt}partial \hspace{-0.4pt}evaluators, \hspace{-0.4pt}and \hspace{-0.4pt}lessons \hspace{-0.4pt}learned.

\end{itemize}

The remainder of this paper is organized as follows. In Sec.~\ref{sec:RTM}, reversible Turing machines are described, and the reversible interpreter is defined, which is used for the inversion by specialization experiment in Sec.~\ref{sec:experiment}. 
In Sec.~\ref{sec:conclusion}, conclusions are drawn, and open points are discussed.

We follow the partial evaluation terminology for {flowchart} languages~\cite{Hatcliff:99:FCL, JGS:93}. Reversible computing from a programming language perspective is presented elsewhere~\cite{GlYo23:perspective}.
In a related experiment~\cite{GKH:03:pepm}, interpreters were transformed into inverse interpreters by partial evaluation of the Universal Resolving Algorithm URA~\cite{AbrGlu02:URAF},
which is an application of the theory of non-standard semantics~\cite{AbrGlu:00:nsint}.

\section{Reversible Turing Machines}
\label{sec:RTM}

\begin{figure}[t] 
\noindent
\centering
(a) \qquad 
\includegraphics[width=0.65\textwidth]{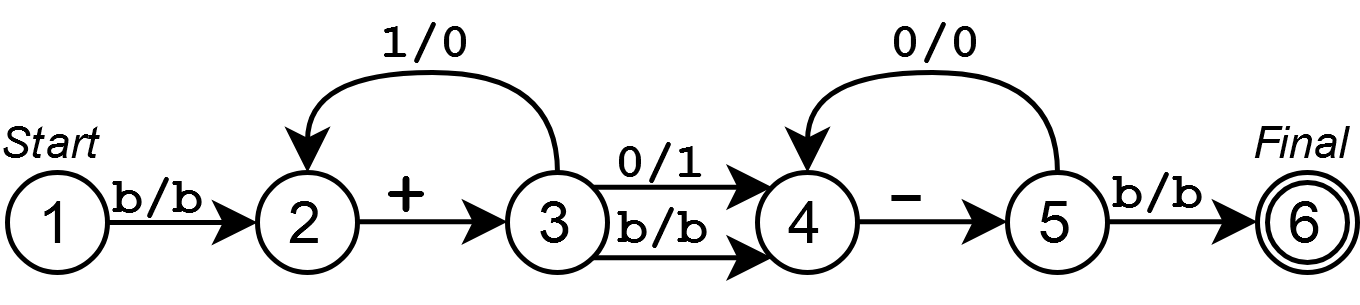}

\vspace{2ex}

(b) \qquad  \includegraphics[width=0.65\textwidth]{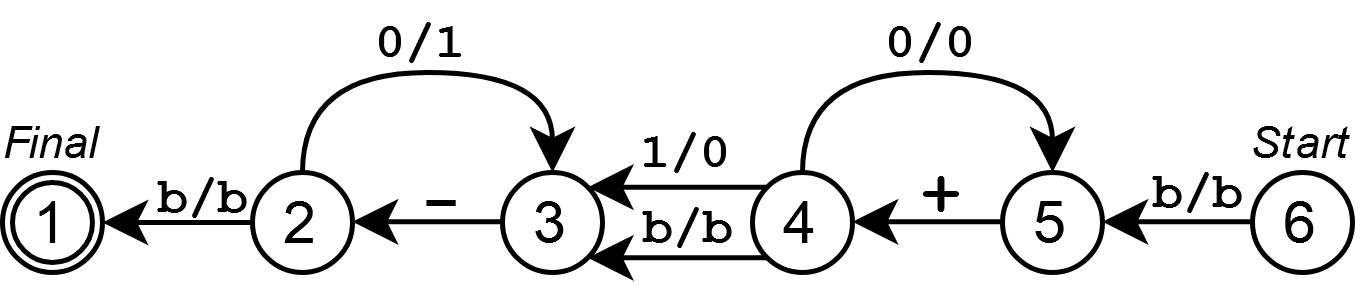}
\caption{Transition diagrams of the RTMs for 
(a) binary increment {\RTMinc} and 
(b) binary decrement {\RTMdec}}
\label{fig:incRTM}
\end{figure}






RTMs 
are forward and backward {deterministic} Turing machines. They operate over a finite set of tape symbols (including the blank symbol $b$) by reading/writing symbols and moving the tape head on a two-way infinite tape. 
An example is the RTM {\RTMinc}, which increments a binary $n$-bit number by 1 (modulo~$2^n$). For simplicity, we assume that the least significant bit comes first. Initially, the tape head is on the left of the first digit. The tape head then shifts to the right, flipping bits until it
flips a ``0'' to a ``1'' or encounters a blank and then returns to
its original position ({\eg}, 1101
\raisebox{0pt}[2.28ex]{$\stackrel{\mbox{\scriptsize\RTMinc}}{\rightarrow}$}
0011, 1111
\raisebox{0pt}[2.28ex]{$\stackrel{\mbox{\scriptsize\RTMinc}}{\rightarrow}$}
0000)~\cite{YAG:08:CF}. 
The transition diagram of {\RTMinc} is shown in Fig.~\ref{fig:incRTM}(a), and the transition rules are shown in Fig.~\ref{fig:incRTMrules}(a).

An RTM is defined by a finite set of symbol and shift 
rules that 
transition from state $q_1$ to state $q_2$ 
of the machine 
when reading/writing 
on the tape 
or moving the tape head:
\begin{itemize}
\item the 
\emph{symbol rule} $(q_1, (s_1,s_2), q_{2})$ 
reads symbol $s_1$ and writes symbol $s_2$ on the tape;
\item the \emph{shift rules}  $(q_1, \mbox{\textbf{--}}, q_2)$ and $(q_1, \mbox{\textbf{+}}, q_2)$  
move the tape head to the left and right, 
respectively.
\end{itemize}

An RTM starts and halts at unique start and final states in a \emph{standard position},
namely, with the tape head on the first blank to the left of a finite blank-free string of tape symbols (the input and output string). 
There must be no transition leading to the start state or out of the final state.
The remaining tape is assumed to contain {an infinite number of} blanks.

A machine is locally \emph{forward} and \emph{backward deterministic} if every state has either (i) one outgoing or incoming shift rule or (ii) one or more outgoing (incoming) symbol rules that read (write) different symbols.
The binary increment machine {\RTMinc} is such an RTM. It is easy to syntactically check whether it is locally forward and backward deterministic by inspecting the rules (either in the diagram or the rule set). 
For example, {state~4 has exactly one} outgoing shift rule and three incoming symbol rules that write three different symbols (0, 1, $b$).
A Turing machine is usually only forward deterministic, whereas an RTM is always forward and backward deterministic. 
RTMs are a proper subset of the Turing machines.

The RTMs were introduced by Bennett~\cite{Bennett:IBM73}; we use the modern \emph{triple transition rules} 
of publication~\cite{AxGl11FoSSaCS}, where a complete  formalization 
is presented. RTMs are the gold standard for reversible programming languages because they compute the \emph{injective computable functions} (for every Turing machine that implements an injective function, there exists an RTM that implements the same function, {and vice versa).} A reversible language that is computationally as powerful as the RTMs is \emph{r-Turing complete}~\cite{AxGl11LATA,AxGl11FoSSaCS}.

Bennett gave a method~\cite{Bennett:IBM73} that can embed any deterministic 
Turing machine in an RTM 
at the cost of computing additional output. 
It is remarkable that every 
irreversible computation can be embedded in a reversible (injective) computation: this is the theoretical basis for 
the possibility of universal, reversible computing devices.
In addition, there are several subclasses, such as the self-inverse RTMs~\cite{Nakano:scico:22}.

\paragraph{Machine Inversion}
\label{sec:RTMinv}

Unlike a conventional Turing machine, the inversion of an RTM into an inverse machine is straightforward: 
\begin{enumerate}
\item swap the unique start and final states, and
\item locally invert each rule as follows:
\begin{tabular}{@{~}l@{~}r@{\,}c@{\,}l@{\,\,to\,\,}r@{\,}c@{\,}l}
   & $(q_1,$ & $(s_1,s_2),$          & $q_2)$ & $(q_2,$ & $(s_2,s_1),$          & $q_1)$,\\
   & $(q_1,$ & $\mbox{\textbf{+}},$  & $q_2)$ & $(q_2,$ & $\mbox{\textbf{--}},$ & $q_1)$,\\
   & $(q_1,$ & $\mbox{\textbf{--}},$ & $q_2)$ & $(q_2,$ & $\mbox{\textbf{+}},$  & $q_1)$.\\
\end{tabular}
\end{enumerate}
The result {of inverting an RTM} is again an RTM. 
An example of this is the inversion of the binary increment machine {\RTMinc} illustrated in Fig.~\ref{fig:incRTM}(a).
The inverse machine {\RTMdec} that decrements a binary number is shown in 
Fig.~\ref{fig:incRTM}(b) and its transition rules in
Fig.~\ref{fig:incRTMrules}(b). 

The inversion of transition diagram~(a) into transition diagram~(b) in Fig.~\ref{fig:incRTM} 
can {also} be performed graphically by swapping the unique start and final states, changing the direction of all arrows, swapping the read/write symbols, and changing the shift direction (\textbf{+} $\leftrightarrow$ \textbf{--}). 
In both cases, the formal and graphical inversion, the result is 
{a} binary decrement machine.
{Note that inversion neither adds nor deletes rules. 
Inverting an RTM twice returns the original RTM.
For example, inverting {\RTMdec} 
gives: %
$({\RTMinc^{-1}})^{-1} = {\RTMinc}$.}

\subsection{Implementation: Reversible RTM Interpreter}
\label{sec:RTMint}

\begin{figure}[t] %
\centering
\begin{minipage}[t]{0.39\textwidth}
(a) Rules of binary increment {\RTMinc}:
\begin{alltt}
((1 . ((BLANK . BLANK) . 2))
 (2 . (RIGHT           . 3))
 (3 . ((0     . 1    ) . 4))
 (3 . ((1     . 0    ) . 2))
 (3 . ((BLANK . BLANK) . 4))
 (4 . (LEFT            . 5))
 (5 . ((0     . 0    ) . 4))
 (5 . ((BLANK . BLANK) . 6)))
\end{alltt}
~~~~~Start state \texttt{1}, final state \texttt{6}.
\end{minipage}
\quad\quad\quad
\begin{minipage}[t]{0.39\textwidth}
(b) Rules of binary decrement {\RTMdec}:
\begin{alltt}
((6 . ((BLANK . BLANK) . 5))
 (4 . ((0     . 0    ) . 5))
 (5 . (RIGHT           . 4))
 (4 . ((BLANK . BLANK) . 3))
 (2 . ((0     . 1    ) . 3))
 (4 . ((1     . 0    ) . 3))
 (3 . (LEFT            . 2))
 (2 . ((BLANK . BLANK) . 1)))
\end{alltt}
~~~~~Start state \texttt{6}, final state \texttt{1}.
\end{minipage}

\vspace{2ex}

\begin{minipage}[t]{0.7\textwidth}
Example input and output tape for {\RTMinc}:~~~~\;$\texttt{(1 1 0 1)} \mapsto \texttt{(0 0 1 1)}$ \newline
Example input and output tape for {\RTMdec}: $\texttt{(0 0 1 1)} \mapsto \texttt{(1 1 0 1)}$
\end{minipage}

\caption{{\RTMinc} and {\RTMdec}: RTM-program representation and example tapes for the RTM-interpreter (Fig.~\ref{fig:rtmcode})}
\label{fig:incRTMrules}
\end{figure}

\begin{figure}[t]
    \centering
    \includegraphics[height=0.55\textheight]{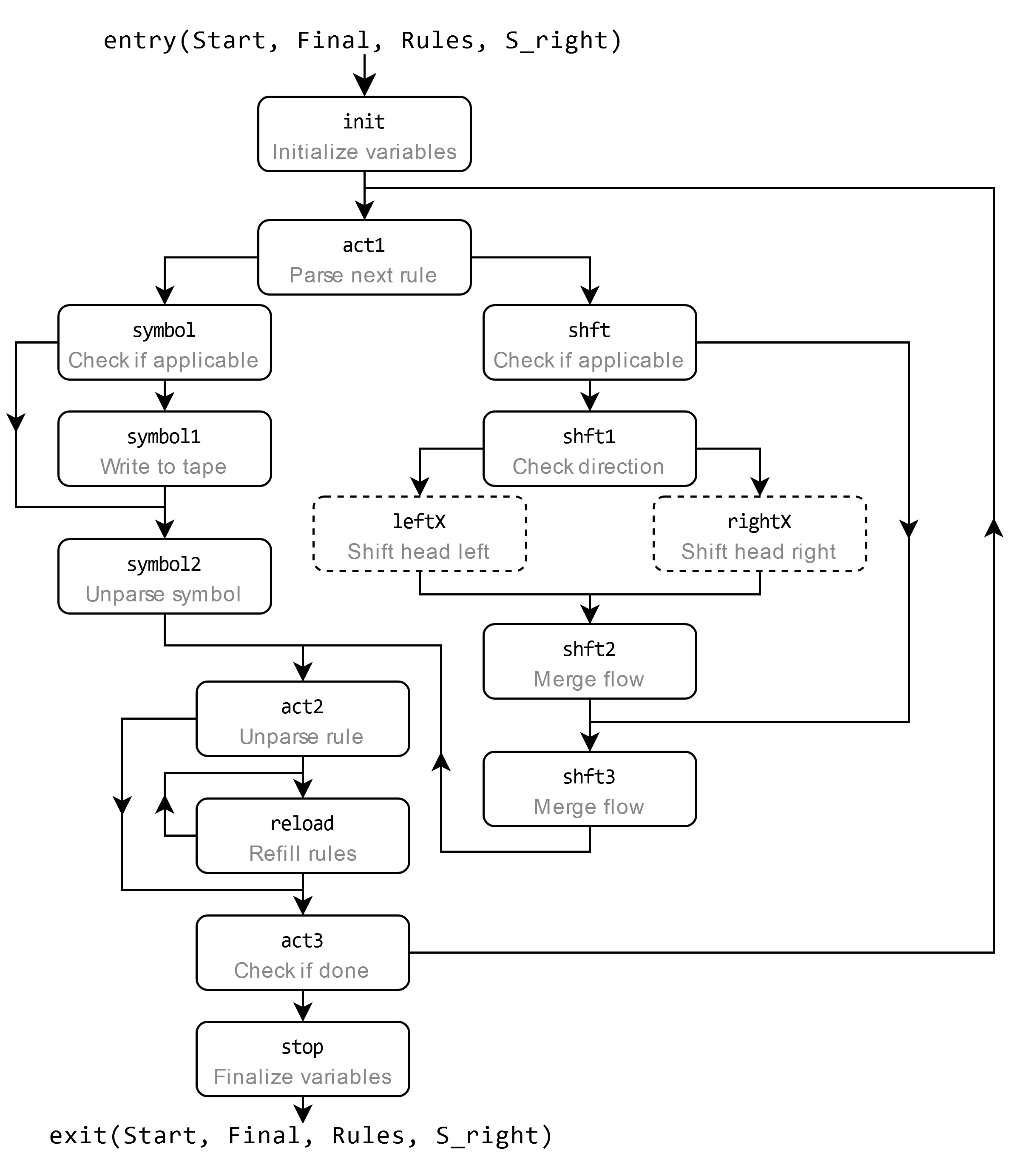}
    \caption{Control-flow diagram
    of the RTM-interpreter}
    \label{fig:rtmflowchart}
\end{figure}

\begin{figure}[p] 
\noindent
\begin{minipage}{0.5\textwidth} 
\begin{lstlisting}[linerange=1-64]
(#Start Final Rules# S_right)
    -> (#Start Final Rules# S_right)
    with (#Q Q1 Q2 S1 S2# S S_left
          #RulesRev Rule R#)
@\codeline@
init: entry
      #Q ^= Start#
      S ^= 'BLANK
      goto act1

act1: #fi !RulesRev && Q = Start#
         #from# init #else# act3
      #(Rule . Rules) <- Rules#
      #(Q1 . (R . Q2)) <- Rule#
      #if R = 'LEFT || R = 'RIGHT#
         #goto# shft #else# symbol

act2: #fi R = 'LEFT || R = 'RIGHT#
         #from# shft3 #else# symbol2
      #Rule <- (Q1 . (R . Q2))#
      #RulesRev <- (Rule . RulesRev)#
      #if Rules goto# act3 #else# reload

reload: #fi Rules# #from# reload #else# act2
        #(Rule . RulesRev) <- RulesRev#
        #Rules <- (Rule . Rules)#
        #if RulesRev# 
           #goto# reload #else# act3

act3: #fi RulesRev#
         #from act2 else reload#
      #if !RulesRev && Q = Final#
         #goto# stop #else# act1

stop: from act3
      S ^= 'BLANK
      #Q	^= Final#
      exit
@\codeline@
symbol: from act1
        #(S1 . S2) <- R#
        if #Q = Q1# && S = #S1#
           goto symbol1 else symbol2

symbol1: from symbol
         #Q ^= Q1#
         #Q ^= Q2#
         S ^= #S1#
         S ^= #S2#
         goto symbol2

symbol2: fi #Q = Q2# && S = #S2#
            from symbol1 else symbol
         #R <- (S1 . S2)#
         goto act2
@\codeline@
shft: from act1
      #if Q = Q1 goto# shft1 #else# shft3

shft1: from shft
       #Q ^= Q1#
       #Q ^= Q2#
       #if R = 'LEFT# 
          #goto# left #else# right
\end{lstlisting} 

\end{minipage}
\hfill 
\begin{minipage}{0.5\textwidth} 
\begin{lstlisting}[linerange=66-999, firstnumber=66]
left: from shft1
      if S_right = 'nil && S = 'BLANK
         goto left_1b else left_1p

left_1b: from left
         S ^= 'BLANK
         goto left1

left_1p: from left
         S_right <- (S . S_right)
         goto left1

left1: fi S_right = 'nil
          from left_1b else left_1p
       if S_left = 'nil
          goto left_2b else left_2p

left_2b: from left1
         S ^= 'BLANK
         goto left2

left_2p: from left1
         (S . S_left) <- S_left
         goto left2

left2: fi S_left = 'nil && S = 'BLANK
          from left_2b else left_2p
       goto shft2
@\codeline@
right: from shft1
       if S_left = 'nil && S = 'BLANK
          goto right_1b else right_1p

right_1b: from right
          S ^= 'BLANK
          goto right1

right_1p: from right
          S_left <- (S . S_left)
          goto right1

right1: fi S_left = 'nil
           from right_1b else right_1p
        if S_right = 'nil
           goto right_2b else right_2p

right_2b: from right1
          S ^= 'BLANK
          goto right2

right_2p: from right1
          (S . S_right) <- S_right
          goto right2

right2: fi S_right = 'nil && S = 'BLANK
           from right_2b else right_2p
        goto shft2
@\codeline@
shft2: #fi R = 'LEFT# 
          #from# left2 #else# right2
       goto shft3

shft3: #fi Q = Q2# #from# shft2 #else# shft
       goto act2
\end{lstlisting}
\end{minipage}
\caption{RTM-interpreter written in {\ARL}. Static operations (in blue) depend only on the static machine description (\texttt{Start}, \texttt{Final}, \texttt{Rules}); all other operations (in black) are residualized by the specializer.}
\label{fig:rtmcode}
\end{figure}

An RTM 
can be run using the reversible RTM-interpreter shown in Fig.~\ref{fig:rtmcode} when represented by a list of rules, as shown in Fig.~\ref{fig:incRTMrules}. The interpreter is written in {\ARL}, 
formerly known as RL when first presented~\cite{NormannGlueck:24}. {\ARL} is an unstructured and reversible flowchart language with symbolic values {and lists as data structures}. 
{The overall structure of the reversible interpreter is shown in the control-flow diagram in Fig.~\ref{fig:rtmflowchart}.}
{The interpreter has a familiar decode-execute loop of a traditional interpreter.}
Notably, 
because an RTM interpreter is implementable in {\ARL}, the language must be r-Turing complete.

The interpreter was {originally} translated from a version written in Janus~\cite{YAG:08:CF} but was later rewritten to be more idiomatic to {\ARL}~\cite{NormannGlueck:24}. 
It %
{uses} the modern interpretation of RTMs~\cite{AxGl11FoSSaCS} 
instead of Bennett's original model {with quadruple rules}~\cite{Bennett:IBM73}.
{Triple rules have the advantage of being symmetric.}

Lines 1-4 in Fig.~\ref{fig:rtmcode} declare the variables {for} the input and output, \texttt{($\ldots$)}~\texttt{->}~\texttt{($\ldots$)}, and those that are only used 
{locally},
\texttt{with}~\texttt{($\ldots$)}. The input 
is the \emph{machine description} (consisting of the start state \texttt{Start}, final state {\texttt{Final}}, and transition rules \texttt{Rules}) as well as the input tape \texttt{S\_right}.
The output is the unchanged machine description and output tape. {For the interpreter to be implementable in a reversible language, it is necessary to return the machine description in the output as well~\cite[Sec.~2]{GlYo16}.}
We assume that all machine {descriptions are} well-formed;
{no error checking is performed by the interpreter}.
{The main program of the interpreter}
comprises a sequence of blocks, each identified using a unique label. The end of a block is always a jump statement, %
as in traditional flowchart programs such as FCL~\cite{Hatcliff:99:FCL}. 
However, because {\ARL} is reversible, it requires a \emph{come-from} statement at the beginning of each block. 
Conditional come-from statements ($\texttt{fi}~e~ \texttt{from}~l_1~\texttt{else}~l_2$) ensure that we can merge two paths in the control flow while ensuring backward determinism.

The program starts in {block} \texttt{init} and 
initializes
\texttt{Q} (the current state) and
\texttt{S}
(the symbol at the current position of the tape head).
The program 
enters the main loop {in the start state},
{iterates} through all rules, and 
{applies} those that match the
{current state and symbol (\texttt{Q}, \texttt{S})} 
until 
the final state {\texttt{Final}} is reached.

The loop itself works by popping off the first rule, \texttt{Rule} (line~13), and decomposing %
it into three components: \texttt{Q1}, \texttt{R}, and \texttt{Q2} (line~14).
{Block} \texttt{act1} branches (line 15) on 
the type of rule (symbol, shift). Once the {rule is handled},
the control flow {merges} again in {block} \texttt{act2}, and the rule {is} pushed onto the list {\texttt{RulesRev}, which} contains the rules {checked} so far. 
If all rules have been {checked}, {that is, \texttt{Rules} is empty} (line 22),
{block \texttt{reload} restores}
the original list in {\texttt{Rules} using \texttt{RulesRev}}. 
Thus, the main loop 
{(blocks \texttt{act1} to \texttt{act3})} 
{continuously iterates} through all rules and applies those that match the current {state and symbol}~(\texttt{Q},~\texttt{S}).

Handling 
symbols rules 
is fairly straightforward {(blocks \texttt{symbol} to \texttt{symbol2})}. The component~\texttt{R} is further parsed to obtain symbols \texttt{S1} and \texttt{S2} of the rule (line~41). If the symbol rule applies (line~42), 
block \texttt{symbol1}
changes the value of \texttt{Q} from \texttt{Q1} to \texttt{Q2} and \texttt{S} from \texttt{S1} to~\texttt{S2}. 

However, shifting left and right {(blocks \texttt{shft} to \texttt{shft3})} is more complicated because of the non-trivial semantics involving the blank symbol. 
{This is necessary to simulate a potentially infinite tape reversibly using two finite lists (\texttt{S\_left}, \texttt{S\_right}).}
The entire right column in Fig.~\ref{fig:rtmcode} is for 
shifting in either direction (lines 66--93 shift left, lines 95--122 shift right). 
{Shifting left (right) is achieved by pushing onto  \texttt{S\_right} (\texttt{S\_left})
and popping an element from 
\texttt{S\_left} (\texttt{S\_right}),
giving a blank if the list is empty. Only a non-blank symbol may be at the end of \texttt{S\_left} and \texttt{S\_right}.
The blocks that push and pop a symbol perform operations that are inverse to each other ({\eg}, blocks \texttt{left} and \texttt{right2}, blocks \texttt{left\_1p} and \texttt{right\_2p}). In a reversible language with procedures that can be called and uncalled, shifting left and right can be implemented by a single procedure~\cite{YAG:08:CF}. The reversible tape simulation %
has been introduced elsewhere~\cite[Sec.~5.2.1]{YAG:08:CF}.
}

\paragraph{Machine Inversion - Revisited} 
\label{sec:RTMinvrevisit}
Although the theoretical 
{description}
of an RTM 
{contains}
an unordered set of rules, this %
abstraction cannot be represented in {\ARL}. 
These rules are represented by an ordered list, as shown in Fig.~\ref{fig:incRTMrules}.
The order %
in which the rules are listed 
{affects} the performance of the interpreter
{because it determines the order in which the main loop checks them}.

{Therefore,} the order 
of rules 
after inversion is important.
There are several ways an RTM inverter can list inverted rules,
all of which are 
{functionally equivalent}
but influence the performance of {the 
inverse machine},
such as leaving the original order unchanged or sorting the rules according to their states.
{However, neither 
preserves the performance {of the original machine}. 
A closer examination of the structure of the interpreter leads to the conclusion that %
the inverter should
\emph{reverse} the original order of the rules instead. 
The inversion shown in Fig.~\ref{fig:incRTMrules}(b) uses this approach}. 

\section{Inversion Projection Experiment}
\label{sec:experiment}
Let us now consider
the experiment illustrated in
Fig.~\ref{fig:experimentsetup}. 
Hereafter,
let $\Lv$ be the language of RTM-programs and $\Rv$ be {\ARL}. {The interpreter}
$\Lv$-$\intpgm_\Rv$ is the RTM-interpreter written in  {\ARL}, 
as shown in Fig.~\ref{fig:rtmcode}.
The inverter $\Lv$-$\inv$ for the RTM-programs was described 
in Sec.~\ref{sec:RTM}. 
The inverter $\Rv$-$\inv$ and specializer $\Rv$-$\spec$ for {\ARL} have been described in a previous study~\cite{NormannGlueck:24}. {Program}
$\pgm_\Lv$ is then {the machine 
description of} 
{\RTMinc}, as shown in Fig.~\ref{fig:incRTMrules}(a). 
{\RTMinc} is {considered} sufficiently nontrivial and, therefore, an exemplary RTM. It has 
all forms of rules, 
states with multiple incoming and outgoing symbol rules, and two loops, as shown in Fig.~\ref{fig:incRTM}(a).

\subsection{Partial Evaluator for RL} %
\label{sec:PEARL}

The partial evaluator for {\ARL} (\emph{\PEARL}) {was} introduced in a previous work~\cite{NormannGlueck:24}.
{\PEARL} is fully implemented in Haskell and is publicly available as part of the complete 
toolset 
for {\ARL}\footnote{\url{https://github.com/Yakokse/PEARL}}. 
It is an offline partial evaluator 
inspired by a three-step procedure from previous partial evaluators of irreversible flowchart languages~\cite{Hatcliff:99:FCL}. However, further action must be taken to ensure that the final residual program is a well-formed reversible program. {\emph{\PEARL} is an offline partial evaluator 
that performs a uniform \emph{binding-time analysis} (BTA) followed by a polyvariant program-point specialization.
For more information on partial evaluation for traditional flowchart languages, see previous publications~\cite{ChristensenGlueck:03:toplas,Hatcliff:99:FCL,JGS:93}.}

The key factors for ensuring well-formed reversible programs have been discussed in detail in previous {publications}~\cite{Mogensen:11,NormannGlueck:24}. 
Similar to how {\ARL}-programs {must always have} a single entry, they must also {have only a} single exit, which means that the {exit} block must 
be specialized only once.
{In some cases}, assertion statements are introduced to {preserve} the {exact} semantics of source programs.
{Static store updates by the partial evaluator} must be injective to generate backward deterministic residual~programs.

\subsection{Results and Discussion}
\label{sec:results}

\begin{figure}[t]
    \centering
    \includegraphics[width=\textwidth]{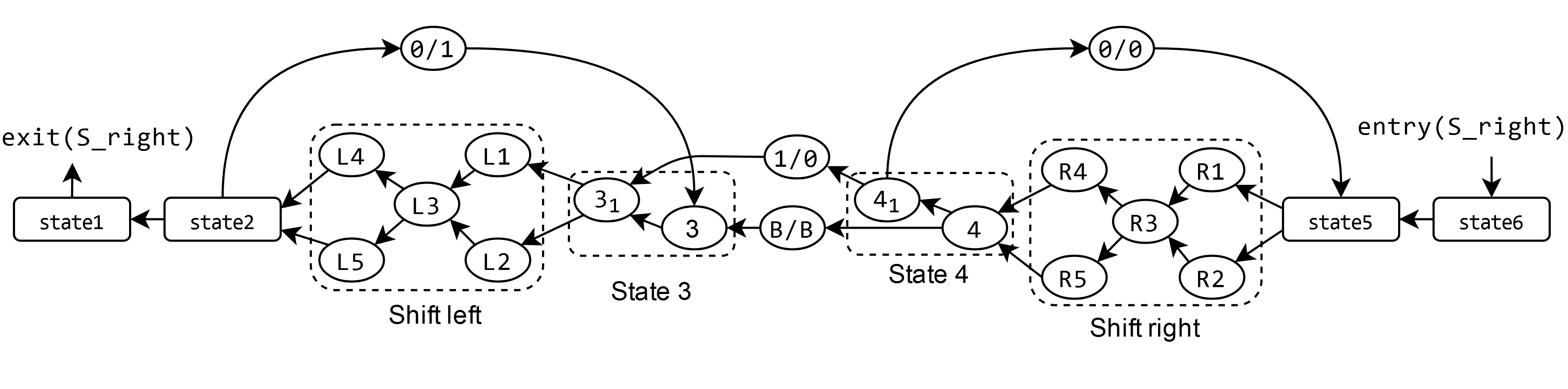}
    \vspace{-4.8ex}
    \caption{Control-flow diagram
    of the residual program  
    corresponding to {\RTMdec}
    in Fig.~\ref{fig:incRTM}(b)}
    \label{fig:residualflowchart}
\end{figure}

\begin{figure}[p] 
\noindent
\begin{minipage}{0.5\textwidth} 
\begin{lstlisting}[linerange=1-64]
(S_right) -> (S_right) with (S S_left)
@\codeline@
state6: entry
        S ^= 'BLANK
        assert(S = 'BLANK)
        S ^= 'BLANK
        S ^= 'BLANK
        assert(S = 'BLANK)
        goto state5
@\codeline@
state5: fi S = '0
           from state4_00 else state6
        if S_left = 'nil 
              && S = 'BLANK
           goto state5_R1
           else state5_R2

state5_R1: from state5
           S ^= 'BLANK
           goto state5_R3

state5_R2: from state5
           S_left <- (S . S_left)
           goto state5_R3

state5_R3: fi S_left = 'nil
              from state5_R1
              else state5_R2
           if S_right = 'nil
              goto state5_R4
              else state5_R5

state5_R4: from state5_R3
           S ^= 'BLANK
           goto state4

state5_R5: from state5_R3
           (S . S_right) <- S_right
           goto state4
@\codeline@
state4: fi S_right = 'nil 
              && S = 'BLANK
           from state5_R4
           else state5_R5
        if S = 'BLANK
           goto state4_BB
           else state4_1

state4_1: from state4
          if S = '1
             goto state4_10
             else state4_00

state4_BB: from state4
           S ^= 'BLANK
           S ^= 'BLANK
           assert(S = 'BLANK)
           goto state3

state4_10: from state4_1
           S ^= '1
           S ^= '0
           goto state3_1
\end{lstlisting}    
\end{minipage}
\hfill 
\begin{minipage}{0.5\textwidth} 
\begin{lstlisting}[linerange=65-999, firstnumber=65]
state4_00: from state4_1
           assert(S = '0)
           S ^= '0
           S ^= '0
           goto state5
@\codeline@
state3: fi S = '1
           from state2_01
           else state4_BB
        goto state3_1

state3_1: fi S = '0
             from state4_10
             else state3
          if S_right = 'nil 
                && S = 'BLANK
             goto state3_L1
             else state3_L2

state3_L1: from state3_1
           S ^= 'BLANK
           goto state3_L3

state3_L2: from state3_1
           S_right <- (S . S_right)
           goto state3_L3

state3_L3: fi S_right = 'nil
              from state3_L1
              else state3_L2
           if S_left = 'nil
              goto state3_L4
              else state3_L5

state3_L4: from state3_L3
           S ^= 'BLANK
           goto state2

state3_L5: from state3_L3
           (S . S_left) <- S_left
           goto state2
@\codeline@
state2: fi S_left = 'nil 
              && S = 'BLANK
           from state3_L4
           else state3_L5
        if S = 'BLANK
           goto state1
           else state2_01

state2_01: from state2
           assert(S = '0)
           S ^= '0
           S ^= '1
           goto state3
@\codeline@        
state1: from state2
        S ^= 'BLANK
        S ^= 'BLANK
        assert(S = 'BLANK)
        S ^= 'BLANK
        exit
@\phantom{\codeline}@ 
\end{lstlisting}    
\end{minipage}    
\caption{Residual {\RTMdec} program in {\ARL} of inversion experiment with {\RTMinc} (reformatted for readability)
}
\label{fig:rtminversionproj}
\end{figure}

We now delve into
the experiment %
described in Fig.~\ref{fig:experimentsetup}, which shows two ways of collapsing the {interpreter tower}. The machine description of the RTM is static  and the input tape is dynamic. We produced two residual programs {with {\PEARL}}:
(i)~the result of specializing the 
interpreter {\wrt} {\RTMdec}, and (ii)~the result of specializing the inverse interpreter {\wrt} {\RTMinc}.

The two residual programs generated by~(i) and~(ii) need not 
be completely 
textually identical to be textually equivalent. A program remains the same {even if}
the blocks are reordered or if the labels and variables are {renamed}.
Not all of these considerations are relevant to this experiment, but {it is necessary to assess} the textual equivalence {of programs} in other {cases}.

When finally compared, the two residual programs were completely identical after relabeling the blocks, that is, (i) and (ii) were textually equivalent.
The overall structure of the residual program is shown in Fig.~\ref{fig:residualflowchart} and the final residual program is shown in Fig.~\ref{fig:rtminversionproj}, with  new labels hinting at the function of each block. 
Blocks responsible for a symbol rule \texttt{X}/\texttt{Y} in the state \texttt{N} are labeled ``\texttt{stateN\_XY}."

The residual program corresponds directly to the transition diagram for {\RTMdec}, as seen in Fig.~\ref{fig:incRTM}(b), as the parsing and rule-matching overhead was eliminated completely. Multiple blocks are required for a single state when it has more than two incoming or outgoing transitions ({\eg}, blocks \texttt{state4} and \texttt{state4\_1}). Path-compression also merges the start state \texttt{state6} with its corresponding symbol rule and similarly for the final state \texttt{state1}.
The blocks required to shift the tape head either left or right (lines 66--93 and lines 95--122, respectively, in Fig.~\ref{fig:rtmcode}) are also fully dynamic and, therefore, preserved in the residual program ({\eg}, 
blocks \texttt{state3\_L1} to \texttt{state3\_L5},
blocks \texttt{state5\_R1} to \texttt{state5\_R5}).

The assertion statements occurring in the residual programs are artifacts of the specialization process; however, in this case, they are redundant
{({\eg}, in block \texttt{state6}).}
{There is a tool} for statically detecting redundant assertions in {\ARL} via abstract interpretation; however, the abstract domain is not sufficiently precise to detect those shown here. Alternatively, we can consider using an automated theorem prover {to eliminate redundant assertions}~\cite{FKG:01:NGC,ReGK:23}
and other optimizations~\cite{DeworetzkiGail:23}.

With these results, the graph in Fig.~\ref{fig:experimentsetup} can be said to truly commute for {\RTMinc}. 
(i) and (ii) are not just the same in the semantic sense, 
but their runtime behavior will also match for all possible input tapes, {as they are textually equivalent}.
We repeated the experiment with other RTMs and confirmed that the residual programs were textually equivalent in all cases.

\paragraph{Reasons for the Successful Projections}

{There are six} degrees of freedom in the experimental setup
in Fig.~\ref{fig:experimentsetup}, which allow for {many subtle variations} that can result in the residual programs not being {textually} equivalent.
Given that we have the {partial evaluator {\PEARL}, the only publicly available specializer for a reversible language, its source language} {\ARL} is an obvious choice for $\Rv$, fixing $\Rv$-$\spec$ and $\Rv$-$\inv$.
{This determines the first three degrees of freedom: $\Rv$, $\Rv$-$\spec$, and $\Rv$-$\inv$.}

{For the remaining three degrees of freedom, $\Lv$, $\Lv$-$\intpgm$, and $\Lv$-$\inv$}, we chose the RTM language for~$\Lv$ owing to its computational power.
However, the 
{exact} 
design and implementation {details} of $\Lv$-$\intpgm$ and $\Lv$-$\inv$ {remained} open {to experimentation}.
This 
resulted in {several} experiments 
{that produced residual programs with minor performance and textual differences}~\cite[Sec.\,5.2]{NormannGlueck:24}.
We have identified many requirements that lead to the successful 
{application of the}
projections.
These are summarized as follows.

\begin{enumerate}
    \item \textit{Alignment of} $\RTM$-$\intpgm$ \textit{{\wrt}~{\PEARL} and} $\ARL$-$\inv$: %
    The dynamic parts of 
    $\RTM$-$\intpgm$
    must be textually equivalent to the {appropriate} dynamic parts of 
    $\RTM$-$\intpgm^{-1}$.
    The parts of 
    $\RTM$-$\intpgm$ that {\PEARL} 
    {determines as}
    dynamic must be symmetric ({\eg}, the blocks for 
    updating and moving the tape head)
    because 
    $\ARL$-$\inv$~\cite[Appdx.\,A]{NormannGlueck:24} 
    {performs inversion purely structurally}.
    However, asymmetric parts 
    are acceptable as long as they are static
    (\eg, block \texttt{reload} {at the end of the main loop}) 
    {because they are evaluated 
    by the partial evaluator
    and are not part of the residual program.}
    
    \item \textit{Alignment of} $\RTM$-$\inv$ \textit{{\wrt}}~$\RTM$-$\intpgm$: 
    Some freedom in 
    $\RTM$-$\inv$ may be restricted by other components. For example, the 
    {reverse}
    order of the inverted rules was determined by how 
    $\RTM$-$\intpgm$
    {traverses the rule list}, as {explained} in Sec.~\ref{sec:RTMint}. 
    If another order had been used, the order of matching symbol rules
    in the residual program 
    could have been different.
    
    \item $\ARL$-$\inv$ and $\RTM$-$\inv$:
    The inverters for the two languages
    were effective in preserving the structure and quality of their respective programs. If they instead generated some artifacts in the code {({\eg}, additional rules)}, it would 
    also possibly 
    affect the residual programs.  
    The machine
    artifacts are directly compiled into the residual program via the Futamura projection. Interpreter artifacts also appear in the residual program if deemed dynamic by the specializer.
    
    \item \textit{\PEARL}: 
    A key characteristics of the uniform BTA used by {\PEARL} is that it is flow-insensitive {({\ie}, a variable has the same binding time at every point in a program).}
    This is a desirable characteristic 
    when we want the
    binding-time annotations of $\RTM$-$\intpgm$ to match those of 
    $\RTM$-$\intpgm^{-1}$.
\end{enumerate}

\paragraph{Generality of Results}

The above discussion shows how interconnected the different parts of the experiment were. An obvious question is whether the 
two residual programs
generated by~(i) and~(ii)
are textually equivalent
for all RTMs and not just for {\RTMinc} {and some other machines}.
Although we have no formal proof, 
let the following argument suffice.

The inversion of 
$\RTM$-$\intpgm$ to $\RTM$-$\intpgm^{-1}$
by $\ARL$-$\inv$ 
is 
performed once
for all RTMs, and
we can examine
the two programs 
independently of a particular RTM. 

The specialization of 
an $\ARL$-program {\wrt}\ static input
follows the program annotations of {\PEARL}'s BTA,
and the outcome of this specialization can be predicted.
This is an advantage of using an offline partial evaluator. 
The BTA only uses the \emph{classification} of
the machine description %
as \emph{static} (\texttt{Start}, \texttt{Final}, \texttt{Rules}) and the input tape as \emph{dynamic} (\texttt{S\_right}). 
It 
guarantees that all operations {that depend only} on the rules ({\eg}, \texttt{Rule}, \texttt{Rules}, and \texttt{RulesRev}) 
{and} the state ({\ie}, \texttt{Q}, \texttt{Q1}, and \texttt{Q2}) are annotated as static (colored blue in Fig.~\ref{fig:rtmcode}) and fully 
{evaluated %
at specialization time}. 
Static operations in $\RTM$-$\intpgm$ and $\RTM$-$\intpgm^{-1}$ are 
thus never part of the residual programs 
({\eg}, the entire blocks \texttt{act1} to \texttt{act3} of the main loop).
Only the operations that depend on the tape (\texttt{S}, \texttt{S\_right}, and \texttt{S\_left}) are annotated as dynamic. 
{These operations occur} in blocks related to moving the tape head and updating the symbol \texttt{S}.
Notably, the annotation of $\RTM$-$\intpgm$ and $\RTM$-$\intpgm^{-1}$ by the BTA is independent of a particular RTM.

Because the same $\RTM$-$\inv$ is used for all RTMs, we know how each rule 
is inverted (Sec.~\ref{sec:RTMinv}) and that the inverted rules are listed in reverse order in the description of the inverse machine 
(Sec.~\ref{sec:RTMinvrevisit}).
It remains to be examined in which order symbol rules are matched, and how the blocks that execute symbol and shift 
rules in $\RTM$-$\intpgm^{-1}$ and $\RTM$-$\intpgm$ are specialized {by {\PEARL}} {\wrt}\ a rule $r$ and its inverse $r^{-1}$.
\begin{itemize}
\item The net effect of the opposite rule orders of an RTM and its inverse is that $\RTM$-$\intpgm^{-1}$ and $\RTM$-$\intpgm$ traverse the 
rule lists 
in the \emph{same order}. 
Thus, if a state has multiple outgoing or incoming symbol rules, they are matched with the dynamic symbol \texttt{S} in the same order in both residual programs. 

\item
By specializing 
the blocks that execute a symbol rule in $\RTM$-$\intpgm$ {\wrt}\ $r^{-1} = (q_2, (s_2,s_1), q_1)$ and by specializing the corresponding inverted blocks in $\RTM$-$\intpgm^{-1}$ {\wrt}\ $r = (q_1, (s_1,s_2), q_2)$, textually equivalent residual blocks are generated, regardless of the 
concrete 
states 
and symbols.

\item
A \emph{left shift} $(q_2, \mbox{\textbf{--}}, q_1)$ in an inverse RTM is implemented in $\RTM$-$\intpgm$ using blocks that move the 
head to the \emph{left}.
A \emph{right shift} $(q_1, \mbox{\textbf{+}}, q_2)$ in the original RTM is implemented in $\RTM$-$\intpgm^{-1}$ using blocks that move the head to the \emph{left}. 
In both cases, the 
residual blocks shift left and are textually equivalent. 
The same applies to a right shift $(q_2, \mbox{\textbf{+}}, q_1)$ and a left shift $(q_1, \mbox{\textbf{--}}, q_2)$.
\end{itemize}
Thus, the order in which the symbol rules are matched is the same in both residual programs, and the residual blocks implementing the 
symbol and shift rules are textually equivalent. This completes the argument for why we conjecture that 
the two residual programs
are %
textually equivalent {for all RTMs.}

\paragraph{Alternative Experiment}

One of the tantalizing 
questions is whether the inversion is neutralized across two levels, {\ie}, whether specializing an inverse interpreter $\intpgm^{-1}$ {\wrt} an inverse program $\pgmq^{-1}$
is as if no inversion was performed at all, neither of %
$\intpgm$ 
nor
$\pgmq$.
That is, whether the 
residual
program is textually equivalent to the residual program generated by specializing $\intpgm$ {\wrt}\ $\pgmq$ according to the first Futamura projection~\cite{Futamura:71}, which corresponds to the translation of $\pgmq_\Lv$ into $\pgmq_\Rv$.

Similar to~(\ref{eq:challenge}), this question can be briefly stated:
\begin{eqnarray}\label{eq:neutralize}
 \sem{\spec}(\intpgm, \pgmq) &\stackrel{\mbox{\footnotesize?}}{\cong}&
 \sem{\spec}(\intpgm^{-1}, \pgmq^{-1})~.
\end{eqnarray}
Again, the residual programs generated by $\spec$ on both sides of~(\ref{eq:neutralize}) 
are 
guaranteed to be 
functionally equivalent~\cite{AbrGlu02:URAF,Futamura:71}, but can they be textually equivalent? 
In our case, where 
$\ARL$-$\inv$ and $\RTM$-$\inv$
perform a structural inversion,
this experiment 
is an instance of the original setup in Fig.~\ref{fig:experimentsetup}.
Everything said
above
also applies to the experimental setup~(\ref{eq:neutralize}),
since $\pgm$ 
can be any RTM and in particular the inverse of an RTM. 
For example, $\RTMdec$ in Fig.~\ref{fig:incRTM} is just another RTM 
of the same 
quality as {\RTMinc}, 
and 
$({\RTMdec})^{-1} = {\RTMinc}$.

Let $\pgm$ be the inverse of an RTM $\pgmq$, that is $\pgm = \pgmq^{-1}$,
then $\pgm^{-1} = (\pgmq^{-1})^{-1} = \pgmq$.  
With the textual equivalence~(\ref{eq:challenge}) established above for our experimental setup, 
the inversion of both levels is always neutralized:
\begin{eqnarray}\label{eq:neutralizeRTM}
 \sem{\PEARL}(\RTM\text{-}\intpgm, \pgmq) &=&
 \sem{\PEARL}(\RTM\text{-}\intpgm, \pgm^{-1})\nonumber\\ &\cong&
 \sem{\PEARL}(\RTM\text{-}\intpgm^{-1}, \pgm)\\ &=&
 \sem{\PEARL}(\RTM\text{-}\intpgm^{-1}, \pgmq^{-1})~.\nonumber
\end{eqnarray}
Also in this case, the residual programs generated for $\pgmq = {\RTMinc}$ and $\pgmq^{-1} = {\RTMdec}$ 
are 
textually equivalent.

In general, %
the textual equivalence~(\ref{eq:challenge}) does not necessarily imply 
the textual equivalence~(\ref{eq:neutralize}). The quality of the inverted programs $\intpgm^{-1}$ and $\pgmq^{-1}$ on the right side of~(\ref{eq:neutralize}) can differ greatly from the quality of 
$\intpgm$ and $\pgmq$ on the left side of~(\ref{eq:neutralize}), so that 
the inversion %
of the two programs 
cannot be neutralized by~$\spec$.

\section{Conclusion and Future Work}
\label{sec:conclusion}

Our goal
was to explore the
interplay 
between partial evaluation and 
program 
inversion in reversible computing. We focused on the computational limit of 
program inversion by interpreter specialization, a possibility
that has been known for some time~\cite{AbrGlu:00:nsint,AbrGlu02:URAF}.
For this purpose, we used an RTM-interpreter because RTMs are as computationally powerful as any reversible programming language~\cite{AxGl11FoSSaCS}. 

We confirmed that the offline partial evaluator for a reversible flowchart language~\cite{NormannGlueck:24} can completely eliminate the interpretive overhead of a reversible interpreter and
that an RTM-interpreter exists for which the first Futamura and inversion projections can produce 
textually equivalent programs.
We found that both methods are equally powerful and that neither projection is fundamentally preferable.
The method that is more effective 
depends, among other things, on the particular interpreter and its source language.
We conclude that both projections must be included in our 
transformation toolbox. 

For irreversible languages, it seems more difficult to achieve the same 
results, regardless of whether inversion is performed at the source or meta-level because the effective inversion of irreversible Turing machines and programming languages remains a challenging {problem (first attempted in the 1950s~\cite{McCarthy:book56}).}
On the other hand, any deterministic Turing machine can be embedded in an RTM 
at the cost of computing additional output~\cite{Bennett:IBM73}. 

Programming languages, in general, and unconventional
program 
transformations, in particular, require
constructive validations.
Although the projections used in this study confirm the theoretical possibility and semantic correctness of 
the transformations, 
only actual 
program constructions can demonstrate the 
real possibility of achieving non-trivial 
results.
The above results 
show the potential 
of combining a small number of basic meta-operations, including specialization and inversion of programs~\cite{GlueckKlimov:94:LMMC}.

The mission continues: to explore strange new worlds of computing!

\paragraph{Acknowledgments} 
We thank Olivier Danvy, Lukas Gail, and Peter Thiemann for their useful comments and
the anonymous reviewers for their constructive feedback on an earlier version of this paper.
It is a pleasure to thank SimCorp for the friendly financial support.


\end{document}